\documentstyle[10pt,epsfig]{article}

\def\bfl{\begin{flushleft}}
\def\efl{\end{flushleft}}
\def\bfr{\begin{flushright}}
\def\efr{\end{flushright}}
\def\bc{\begin{center}}
\def\ec{\end{center}}
\def\be{\begin{equation}}
\def\ee{\end{equation}}
\def\ba{\begin{eqnarray}}
\def\ea{\end{eqnarray}}
\def\nn{\nonumber }
\def\lb#1{\label{#1}}

\def\text#1{\mbox{#1}}
\def\drm{d}
\def\schrod{Schr\"oedinger  }

\def\Sign#1{\, \text{sgn}\left(#1\right) }
\def\Der#1#2{\,\frac{\partial #1}{\partial #2}}
\def\Lim#1#2{\, \lim\limits_{#1 \to #2} }

\def\JacobiSn#1#2#3{\, \text{sn}^{#1}\left(#2,\,#3 \right) } 
 
\def\Sech#1#2{\, \text{sch}^{#1}\left(#2 \right) }
\def\HypGF#1#2#3#4{\,\text{F}\left(#1,\,#2,\,#3;\,#4 \right) } 

\begin{document}

~\\
~\\
~\\
~\\
\bfr
Phys. Lett. B 450 (1999) 397-404\\
\efr
\bc
{\LARGE \bf
Nonminimal particle-like solutions in cubic scalar field theory
}

~~\\
{\large 
Konstantin G. Zloshchastiev\\

~\\
E-mail: zlosh@email.com.\\
URL(s): http://zloshchastiev.webjump.com, http://zloshchastiev.cjb.net
}\\
~\\
Received 19 December 1998

\ec

~~\\

\abstract{
The cubic scalar field theory admits the 
bell-shaped solitary wave solutions which can be interpreted as  
massive Bose particles.
We rule out the nonminimal p-brane action for such a solution as the 
point particle with curvature.
When quantizing it as the theory with higher derivatives,
it is shown that the corresponding quantum equation has SU(2) 
dynamical symmetry group
realizing the exact spin-coordinate correspondence.
Finally, we calculate the quantum corrections to the 
mass of the bell boson which can not be obtained by means
of the perturbation theory starting from the vacuum sector.
}

~\\

PACS number(s):  11.10.Lm, 11.15.Ex, 11.27.+d, 11.30.Na\\

~~\\

\large

\section{Introduction}\lb{s-i}

The kink solutions are known to be the solitary wave solutions appearing 
in the relativistic $\varphi^4$ or sine-Gordon models and admitting the 
interpretation in terms of a particle.
Their classical and quantum properties are already studied in a lot of
works \cite{dhn,kpp,col}.
The aim of this paper is to study another, bell-shaped, solitary wave 
solution which has to be the only particle-like solution in the 
$\varphi^3$ theory.
The approach developed in \cite{kpp} consists in the 
constructing of the p-brane action where the nonminimal
terms depending on the world-volume curvature are induced
by the field fluctuations in the neighborhood of the static solution.
When requiring these fluctuations to be damping at infinity,
the effective action evidently arises after nonlinear reparametrization
of the initial theory and excluding of zero field oscillations.

The paper is arranged as follows.
In Sec. \ref{s-kb} we obtain $\varphi^3$-bell solution and study 
its properties on the classical level.
In Sec. \ref{s-ea} we perform the nonlinear parametrization of the 
$\varphi^3$ action by means of the Bogolyubov transition to the 
collective degrees of freedom.
After this, minimizing the action with respect to field fluctuations,
we remove zero modes and obtain the effective action containing
the curvature terms. 
Sec. \ref{s-q} is devoted to quantization of this action as 
the constrained theory with higher derivatives.
In result we obtain the \schrod wave equation describing
wave functions and mass spectrum of the quantum bell boson.
Then we calculate the zeroth and first excited levels to rule out the
bell particle mass with quantum corrections.
Conclusions are made in Sec. \ref{s-c}.

\section{Particle-like solution}\lb{s-kb}

Let us consider the (1+1)-dimensional action
\begin{equation}
S[\phi] = \int  L(\varphi)\, \drm^2 x,              \lb{eq1} 
\end{equation}
\be                                                        \lb{eq2}
L(\varphi) = \frac{1}{2} (\partial_m \varphi) (\partial^m \varphi)  - 
\lambda \varphi  
\biggl( \varphi - \frac{m^2}{\lambda} \biggr)^2 + 
\frac{2}{\lambda^2}
\left(
      \frac{m^2}{3}
\right)^3
\left(1-\Sign{m^2}\right),                                      
\ee
where $\varphi(x,t)$ is the dimensionless scalar field, 
$\lambda \in \Re$.
The last term in eq. (\ref{eq2}) is introduced in such a way that 
the potential energy 
\be                                                        \lb{eq3}
U (\varphi) = 
\lambda \varphi  
\biggl( \varphi - \frac{m^2}{\lambda} \biggr)^2 -
\frac{2}{\lambda^2}
\left(
      \frac{m^2}{3}
\right)^3
\left(   1-\Sign{m^2}
\right)                                      
\ee
would approach zero in the appropriate local minimum point.
For definiteness below we will assume $m \in \Re$ hence
\be                                                        \lb{eq4}
L(\varphi) = \frac{1}{2} (\partial_m \varphi) (\partial^m \varphi)  - 
\lambda \varphi  
\biggl( \varphi - \frac{m^2}{\lambda} \biggr)^2,                                      
\ee
and the system has a single local minimum point.
Therefore, the state $\varphi= m^2/\lambda$ 
has to be (locally) the most energetically favorable, and
stability of it grows as the barrier's height, 
\[
\Delta U =
\frac{1}{3} 
\left(
\frac{2 m^3}{3 \lambda}
\right)^2,
\]
increases.
We will find solutions of the corresponding equation of motion,
\be                                                           \lb{eq5}
\partial^m \partial_m \varphi + 
\lambda (\varphi - m^2/\lambda) (3 \varphi - m^2/\lambda) = 0,
\ee
in the class of solitary waves
\be                                                           \lb{eq6}
\varphi(\rho) = \varphi
\left(
\frac{x-v t}{\sqrt{1-v^2}}
\right),
\ee
hence
\be                                                           \lb{eq7}
\varphi_{\rho\rho} - 
\lambda (\varphi - m^2/\lambda) (3 \varphi - m^2/\lambda) = 0.
\ee
The general integral of this equation
can be expressed in terms of the elliptic functions
\be                                                           \lb{eq8}
\frac{\varphi-m^2/\lambda - \alpha_3}{\alpha_2- \alpha_3} = 
\JacobiSn{2}
         {
          \sqrt{
                \frac{-\lambda}{2\text{k}} 
               }
          \rho
         }
         {\text{k}},
\ee
where $\rho_0$ is supposed zero,
\[
\text{k} =  \frac{\alpha_2 - \alpha_3}{\alpha_1 - \alpha_3}, 
\]
$\alpha_i$ are roots of the cubic equation
\be                                                           \lb{eq9}
\alpha^3 + \frac{m^2}{\lambda} \alpha^2 + C = 0,
\ee
and $C$ is an integration constant.

Among the solutions (\ref{eq8}) we are needed in the regular solitary waves 
with the localized energy density
\be                                                           \lb{eq10}
\varepsilon (x,t) = 
\Der{L}{(\partial_0 \varphi)} \partial_0 \varphi - L.
\ee

It can be checked immediately that the only such a solution
is the bell-shaped one:
\be                                                           \lb{eq11}
\varphi_b (\rho) = 
\frac{m^2}{\lambda} 
\tanh{\!^2 
          \left(\frac{m \rho}{\sqrt{2}}\right)
     },
\ee
having the energy density
\be                                                         \lb{eq12}
\varepsilon_b (x,t) = \frac{2 (m^3/\lambda)^2}{1-v^2}
\Sech{4}
     {
      \frac{m\rho}{\sqrt{2}}
     }
\left[
1-    
\Sech{2}
     {
      \frac{m\rho}{\sqrt{2}}
     }
\right].
\ee
Therefore, it can be interpreted as the relativistic particle with the
energy
\be                                                            \lb{eq13}
E_{\text{class}} = \int\limits_{-\infty}^{+\infty}
\varepsilon_b (x,t)\ \drm x =
\frac{\mu}{\sqrt{1-v^2}},
\ee
where the classical mass of the bell-particle is 
\be                                                            \lb{eq14}
\mu = \frac{8 \sqrt{2}}{15} \frac{m^5}{\lambda^2}.
\ee

\begin{figure}
\centerline{
\epsfysize=.55\textwidth
\epsfbox{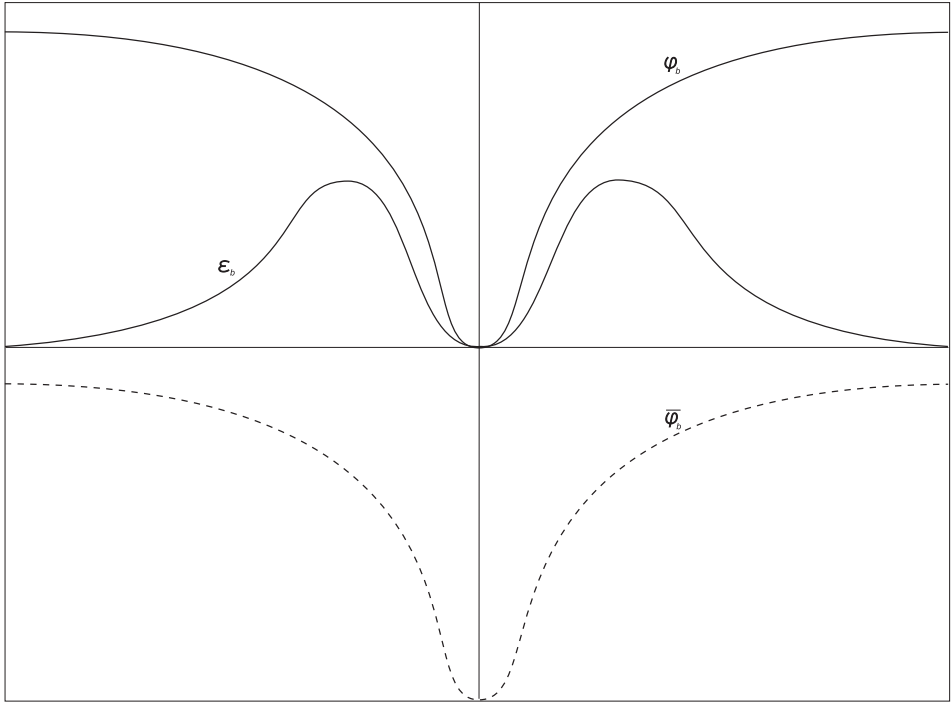}\\}
 \caption{The qualitative plot of the potentials and energy density
for the bell (solid line) and anti-bell (broken line) field solutions}
\label{fig1}
\end{figure}

One should done the following remarks upon the features of the 
solution (\ref{eq11}).
At first, it is topologically trivial unlike, e.g., the kink solutions
in the $\varphi^4$ or sin-Gordon theories.
It could give rise to certain problems with the stability against small
perturbations but following the theorem proven below the 
corresponding 
effective p-brane action for the bell boson appears to be stable.
Second, unlike the $\varphi^4$ or sin-Gordon cases 
the function $-\varphi_b$ has not to be the solution of (\ref{eq7}).
It means that the particle-antiparticle interpretation in the 
$\varphi^3$ theory differs from that in the $\varphi^4$ one and should 
be modified.
One can show that the model (\ref{eq2}) at $m^2 < 0$ also admits the 
single particle-like solution
\be                                                           \lb{eq15}
\bar\varphi_b (\rho) = 
\frac{m^2}{\lambda} 
\left[
\Sech{2}{ 
         \frac{|m| \rho}{\sqrt{2}}
        }
+ \frac{1}{3}
\right],
\ee
having the same energy (\ref{eq12}) - (\ref{eq14}).
From the figure \ref{fig1} one can see that the bell and anti-bell solitons
occupy the different level lines shifted with respect to each other.
It evidently has to be the consequence of the fact that the symmetry 
$\varphi \to - \varphi$ in the cubic SFT is broken initially.

\section{Effective action}\lb{s-ea}

In this section we will construct the nonlinear effective action of 
the $\varphi^3$ theory (\ref{eq1}), (\ref{eq4}) 
about the static bell solution (\ref{eq11}).
Let us introduce the set of the collective coordinates 
$\{\sigma_0=s,\ \sigma_1=\rho\}$ such that
\be                                                      
x^m = x^m(s) + e^m_{(1)}(s) \rho,\ \              
\varphi(x,t) = \widetilde \varphi (\sigma),     \lb{eq16}
\ee
where $x^m(s)$ turn to be the coordinates of a (1+1)-dimensional point
particle, $e^m_{(1)}(s)$ is the unit spacelike vector orthogonal
to the world line, hence
the components of the Frenet basis are
\[
e^m_{(0)} = \frac{\dot x^m}{\sqrt{\dot x^2}},\
e^m_{(1)} = - \frac{1}{\sqrt{\dot x^2}} \frac{\dot e^m_{(0)}}{k},
\]
where the dot means the derivative with respect to $s$.
Then the action (\ref{eq1}) can be rewritten in the
new coordinates as
\be                                                         \lb{eq17}
S[\widetilde \varphi] = 
\int L (\widetilde \varphi) \,\Delta \ \drm^2 \sigma,
\ee
where
\[
\Delta = \text{det} 
\left|
\left|
      \Der{x^m}{\sigma^k}
\right|
\right|
= \sqrt{\dot x^2} (1- \rho k),
\]
\[
L (\widetilde \varphi) = \frac{1}{2}
\left[
      \frac{(\partial_s \widetilde\varphi)^2}{\Delta^2} - 
                  (\partial_\rho \widetilde\varphi)^2
\right]- 
\lambda \varphi  
\biggl( \varphi - \frac{m^2}{\lambda} \biggr)^2,                                      
\]
and $k$ is the curvature of a particle world line
\be                                                            \lb{eq18}
k = \frac{\varepsilon_{mn} \dot x^m \ddot x^n}{(\sqrt{\dot x^2})^3},
\ee
where $\varepsilon_{m n}$ is the unit antisymmetric tensor.
This new action contains the redundant degree of freedom which eventually
leads to appearance of the so-called ``zero modes'' \cite{raj}.
To eliminate this nonphysical 
degree of freedom we must constrain the model
by means of the conditions of both
the vanishing of the functional derivative 
with respect to field fluctuations about some chosen static solutions 
and damping of the fluctuations at infinity,
and in result we will obtain the required effective action.

So, the fluctuations of the field $\widetilde\varphi (\sigma)$ in the 
neighborhood of some static solution $\varphi_{\text{st}} (\rho)$
are given by the expression
\be                                               
\widetilde\varphi (\sigma) = 
\varphi_{\text{st}} (\rho)  + \delta\widetilde\varphi (\sigma).
\ee
Substituting it into the action (\ref{eq17}) and considering the static
equation of motion (\ref{eq4}) for $\varphi_{\text{st}}$ we have
\begin{eqnarray}
S[\delta \widetilde{\varphi}] 
&=& \int d^2 \sigma \ 
   \Biggl\{\Delta 
           \Biggl[ L(\varphi_{\text{st}}) + 
                  \frac{
                        \left( 
                              \partial_s \ \delta \widetilde{\varphi} 
                        \right)^2 
                       }
                       {2 \Delta^2} 
                  - \frac{1}{2} 
                  \Bigl( 
                        \partial_{\rho}  \delta \widetilde{\varphi} 
                  \Bigr)^2 - \nonumber \\
&&                \lambda
                  \biggl( 
                         3 \varphi_{\text{st}} - 2\frac{m^2}{\lambda} 
                  \biggr) 
                  \delta \widetilde{\varphi}^2
           \Biggr]                                             
           + \varphi_{\text{st}}^\prime \delta \widetilde{\varphi} 
           \partial_{\rho} \Delta  + O (\delta \widetilde{\varphi}^3)
    \Biggr\},                                            \nn
\ea
\be                                                       \lb{eq20}
L(\varphi_{\text{st}}) = 
 - \frac{1}{2} (\varphi_{\text{st}}^\prime)^2  - 
\lambda \varphi_{\text{st}}  
\biggl( \varphi_{\text{st}} - \frac{m^2}{\lambda} \biggr)^2,                                      
\ee
where the prime means the derivative with respect to $\rho$.
Extremalizing this action with respect to 
$\delta \widetilde{\varphi}$ one can obtain 
the equation in partial derivatives for field fluctuations:
\[
\Biggl[ \partial_s \Delta^{-1} \partial_s -
     \partial_{\rho} \Delta \partial_{\rho} +
     2 \lambda \Delta 
     \left( 
           3 \varphi_{\text{st}} - 2 \frac{m^2}{\lambda}
     \right) 
\Biggr] \delta \widetilde{\varphi} 
+ \varphi_{\text{st}}' k\sqrt{\dot{x}^2} =
O(\delta \widetilde{\varphi}^2),      
\]
which when substituting 
\[
\varphi_{\text{st}} = \varphi_b, \,
\rho=\epsilon u, \,
\delta \widetilde\varphi = \frac{m^2}{\lambda} \delta X,
\] 
in the linear approximation has the form: 
\be                                                          \lb{eq21}
\Biggl[\epsilon^2 \partial_s \Delta^{-1} \partial_s -
     \partial_u \Delta \partial_u +
     4 \Delta 
     \left( 3 X_0 - 2 \right) 
\Biggr] \delta X 
+  \epsilon X_0^\prime k\sqrt{\dot{x}^2} = 0,      
\ee
where
\[
\epsilon = \frac{\sqrt{2}}{m},\,
X_0 = \text{tanh}^{2} u, \,
\Delta = \sqrt{\dot x^2} (1-\epsilon u k).
\]

To resolve this very complicated equation we suppose 
\be                                                 \lb{eq22}
\delta X (s,u) = \epsilon k(s) f(u),
\ee
and, expanding eq. (\ref{eq21}) in the Taylor series with respect to 
$\epsilon$ in such a way that
\[
\Lim{\epsilon}{0} \epsilon^2 \frac{1}{\sqrt{\dot{x}^2}} \frac{d}{ds} 
\frac{1}{\sqrt{\dot{x}^2}} \frac{dk}{ds} \not= 0,    
\]
we have in the linear approximation the partitioned system of the two 
ordinary differential equations
\be  
{\epsilon}^2 \frac{1}{\sqrt{\dot{x}^2}} \frac{d}{ds} 
\frac{1}{\sqrt{\dot{x}^2}} \frac{dk}{ds} +ck = 0,    \lb{eq23}
\ee
\be
-f_{uu} + 4
\left( 3X_0 - 2 \right) f 
- cf + X'_0 = 0.       \lb{eq24}
\ee 

First of all, we are needed to find the solution of the last equation
such that field fluctuations vanish at both infinities, i.e.,
we suppose the next boundary conditions:
\be                                                   \lb{eq25}
f(+\infty) = f(-\infty) =0,
\ee 
evidently corresponding to the singular Stourm-Liouville problem 
which can be completely resolved in our case.

{\it Theorem.}
The differential equation (\ref{eq24}) 
has the two eigenfunctions and eigenvalues 
satisfying with the boundary conditions (\ref{eq25}):
\ba
&&f_+ = \frac{C_+ (\text{cosh}^2 u - 5/4) + 2 \sinh u}
             {3 \text{cosh}^3 u},\ \ c_+ = 3,                     \nn\\
&&f_- = \frac{C_- - 2 \sinh u }
             {5 \text{cosh}^3 u},\ \ c_- = -5,                    \nn
\ea
where $B_\pm$ are arbitrary integration constants.

{\it Proof.}
Firstly we consider the case $c=0$.
Then the general integral of eq. (\ref{eq24}) is
\be                                               \lb{eq26}
f_{c=0} = 
\frac{\tanh u}{15} 
\left[
      3 + \text{cosh}^2 u  + \frac{C_1}{\text{cosh}^2 u}
\right]
+ 
C_2 \text{cosh}^4 u\,
    \HypGF{1}{3}
          {\frac{9}{2}}{\text{cosh}^2 u},
\ee
where $\HypGF{a}{b}{c}{z}$ is the hypergeometric function, $C_i$ are
integration constants.
As one can see this solution does not satisfy with (\ref{eq25}).
Therefore, below we will assume $c\not=0$.

Performing the variable change $z = \text{cosh}^2 u$, 
we can rewrite (\ref{eq24}) in the form
\be                                                \lb{eq27}
2 z (z-1)  f_{z z} + (2 z -1)  f_{z} - 2
\left(1 - \frac{ c}{4} - \frac{3}{z}
\right)  f = \frac{\sqrt{z-1}}{z^{3/2}},
\ee
hence after the shifting substitution
\[
f = \widetilde f +  \frac{2}{c} \frac{\sqrt{z-1}}{z^{3/2}}
\]
we obtain the homogeneous equation
\be                                                 \lb{eq28}
2 z (z-1) \widetilde f_{z z} + (2 z -1) \widetilde f_{z} - 2
\left(
      \widetilde c^2 - \frac{3}{z}
\right) 
\widetilde f = 0,
\ee
where
\be                                                \lb{eq29}
c = 4 (1- \widetilde c^2),\ \ 
|\widetilde c| \not= 1.
\ee
Then the bound state condition (\ref{eq25}) should be rewritten as
\be                                                         \lb{eq30}
\widetilde f (1) =0,\ \widetilde f (+\infty) =0.
\ee
The general integral of eq. (\ref{eq28}) 
can be expressed in the form:
\[
\widetilde f = 
\frac{C_1}{z^{3/2}}
\HypGF{-\frac{3}{2} - \widetilde c}
      {-\frac{3}{2} + \widetilde c}
      {-\frac{5}{2}}
      {z}
+ C_2 z^2
\HypGF{2 - \widetilde c}
      {2 + \widetilde c}
      {\frac{9}{2}}{z}.
\]
Expanding the hypergeometric functions in series in the neighborhood of
$z=1$, it is straightforward to check that the first from the conditions
(\ref{eq30}) will be satisfied if we suppose
\[
C_2 = \frac{8 \widetilde c}{1575} 
(\widetilde c^2 - 1)
(4\widetilde c^2 - 1)
(4\widetilde c^2 - 9)
\tan{(\pi \widetilde c)} C_1 \equiv - C^{(\text{reg})} C_1,
\]
hence 
\[
|\widetilde c| \not= n+1/2 \ \ \text{except} \ \ 
|\widetilde c| = 1/2, 3/2.
\]

Therefore, the solution of eq. (\ref{eq29}) vanishing at $z=1$ is the 
function
\be                                                         \lb{eq31}
\frac{1}{C_1} \widetilde f^{(\text{reg})} = 
z^{-3/2}
\HypGF{-\frac{3}{2} - \widetilde c}
      {-\frac{3}{2} + \widetilde c}
      {-\frac{5}{2}}
      {z}
- C^{(\text{reg})} z^2
\HypGF{2 - \widetilde c}
      {2 + \widetilde c}
      {\frac{9}{2}}{z}.
\ee
To specify the parameters at which $f^{(\text{reg})}$ would satisfy
with the second condition (\ref{eq30}) we should consider the 
asymptotical behavior of $\widetilde f^{(\text{reg})}$ at large $z$.
We have
\be                                                            \lb{eq32}
\frac{1}{C_1} \widetilde f^{(\text{reg})} (z \to \infty) =
\frac{\text{sin}^2 (2 \pi \widetilde c)}
     {4 \pi^{7/2}}
\left[
       A (\widetilde c, z) + A (-\widetilde c, z)
\right],
\ee
where
\ba
&&A (a, x) = 
\Gamma (2 a)
\biggl[
      \frac{8 i}{15} \Gamma (a+2) \Gamma (a+5/2) +  \nn\\
&& \frac{315}{48} C^{(\text{reg})} \Gamma (a - 3/2) \Gamma (a-1)
\biggr]
(-x)^a 
\left( 1 + O (1/x) \right). \nn
\ea
From this expression and eqs. (\ref{eq29}) and (\ref{eq31}) it can 
easily be seen that 
$\widetilde f^{(\text{reg})}$ does not satisfy with the Stourm-Liouville
conditions everywhere except the points:
\[
|\widetilde c| =  1/2, 3/2, 
\]
or, following (\ref{eq29}),
\[
c = 3, -5,
\]
respectively.
The corresponding eigenfunctions can be obtained directly from 
(\ref{eq31}), Q.E.D.

By virtue of this theorem we have recently obtained all the 
necessary functions to construct the 
effective action for the $\varphi^3$ theory about the 
static bell solution.
Taking into account eqs. (\ref{eq21}) and (\ref{eq22}) and the theorem,
the action (\ref{eq20}) can be rewritten in the explicit p-brane form
\be                                                              \lb{eq33}
S_{\text{eff}} = 
S_{\text{eff}}^{\text{(class)}} + S_{\text{eff}}^{\text{(fluct)}} =
- \int \drm s \sqrt{\dot x^2} 
\left(
       \mu + \alpha_\pm k^2
\right),
\ee
where
\[
\mu = - \int\limits_{-\infty}^{+\infty}   (1 - \rho k)
L (\varphi_b) \drm \rho \mapsto
\int\limits_{-\infty}^{+\infty} \varepsilon_b (\rho)\  \drm \rho \Bigr|_{v=0},
\]
see eq. (\ref{eq14}), and
\be                                                 \lb{eq34}
\alpha_\pm = \frac{\epsilon}{2} 
\left(
      \frac{m^2}{\lambda}
\right)^2
\int\limits_{-\infty}^{+\infty} f_\pm X'_0\  \drm u =
\frac{\mu}{c_\pm m^2}.
\ee
Thus, one can see that the fluctuational corrections lead to appearance
of two
different $\alpha$ (unlike the single $\alpha$ in \cite{kpp}), 
therefore, we have the
bifurcation of the bell solution (\ref{eq11})
as the particle with curvature.
However, in the following section it will be shown that on the 
quantum level nonminimal term, leading to the quantum mass corrections, 
is required to be such that $c>0$.

Finally, the action (\ref{eq33}) yields the equations of motion for the 
bell field solution as a p-brane:
\be
\epsilon^2 
\frac{1}{\sqrt{\dot x^2}}
\frac{\drm}{\drm s}
\frac{1}{\sqrt{\dot x^2}}
\frac{\drm k}{\drm s} +
\left(
      c_\pm - \frac{\epsilon^2}{2} k^2
\right) k = 0.
\ee
Considering the proven theorem, one can see that 
eq. (\ref{eq23}) was just the linearized version of this expression.

\section{Quantization}\lb{s-q}

In the previous section we obtained classical 
effective actions for the model in question.
Thus, to quantize them we must consecutively construct the 
Hamiltonian structure
of the point particle with curvature.
From eqs. (\ref{eq18}) and (\ref{eq33}) one can see that we have the 
theory with higher derivatives \cite{dhot}.
Hence, below we will treat the coordinates and momenta as the 
canonically independent coordinates of phase space.
The phase space consists of the two pairs of 
canonical variables:
\ba
&&x_m,\ \ p_m = \Der{L_{\text{eff}}}{q^m} - \dot \Pi_m, \\
&&q_m = \dot x_m,\ \ \Pi_m =\Der{L_{\text{eff}}}{\dot q^m},
\ea
hence we have
\ba
&&p^n = - e^n_{(0)} \mu 
\left[
      1- \frac{1}{c_\pm m^2} k^2
\right] +
\frac{2 \mu}{c_\pm m^2}
\frac{e^n_{(1)} }{\sqrt{q^2}} \dot k,  \\
&&
\Pi^n = - \frac{2 \mu}{c_\pm m^2} 
\frac{e^n_{(1)}}{\sqrt{q^2}} k.                                \lb{eq39}
\ea
Besides, the Hessian matrix constructed 
from the derivatives with respect to accelerations
appears to be singular that points out the presence of the
constraints on the phase variables of the theory.
There exist the two primary constraints of first kind
\ba
&&\Phi_1 = \Pi^m q_m \approx 0, \\
&&\Phi_2 = p^m q_m +  \sqrt{q^2} 
\left[
      \mu + \frac{c_\pm m^2}{4 \mu} q^2 \Pi^2
\right]  \approx 0,
\ea
besides we should add the proper time gauge condition,
\be
G = \sqrt{q^2} - 1 \approx 0,
\ee
to remove the non-physical gauge degree of freedom.
Then, when introducing the new variables,
\be
\rho = \sqrt{q^2},\ \ v = 
\text{arctanh} 
\left(
      p_{(1)}/p_{(0)}
\right),
\ee
the constraints can be rewritten in the form
\ba
&&\Phi_1 = \rho \Pi_\rho, \nn\\
&&\Phi_2 = \rho 
\left[
      -\sqrt{p^2} \cosh{v} + \mu -
      \frac{c_\pm m^2}{4\mu}
      \left( 
            \Pi^2_v - \rho^2 \Pi^2_\rho
      \right)
\right],                                       \\
&&G=\rho-1, \nn
\ea
hence finally we obtain the constraint
\be
\Phi_2 = 
      -\sqrt{p^2} \cosh{v} + \mu -
      \frac{c_\pm m^2}{4\mu} \Pi^2_v \approx 0,                                       
\ee
which in the quantum theory ($\Pi_v = - i \partial/\partial v$) 
yields 
\[
\widehat\Phi_2 |\Psi\rangle =0.
\]
As was shown by Kapustnikov {\it et al}, the constraint $\Phi_2$ on the 
quantum level admits several coordinate representations that,
generally speaking, lead to different nonequivalent theories, therefore,
the choice between the different forms of 
$\widehat\Phi_2$ should be based on the physical relevance.
Then the physically admissible
equation determining quantum dynamics of the quantum
bell particle has the form:
\be                                                            \lb{eq46}
[ \widehat H-\varepsilon] \Psi(\zeta) = 0, 
\ee 
\be
\widehat H =  -\frac{\drm^2}{\drm \zeta^2} +
  \frac{B^2}{4}
  \sinh{\! ^2 \zeta}
  -B
  \left(
        S+\frac{1}{2}
  \right)
  \cosh{\zeta},                                             
\ee
where $S=0$ in our case, and
\ba
&&\zeta=v/2,\ \sqrt{p^2} = M,                           \nn\\
&&B= \frac{8 \sqrt{2}}{m}
     \sqrt{
           \frac{\mu M}{c_\pm}
          },                                              \lb{eq48}\\
&& \varepsilon =
\frac{16 \mu^2}{c_\pm m^2} 
\left(
      1 - \frac{M}{\mu}
\right). \nn
\ea
It can readily be seen that for the case $c_-$ the ratio $M/\mu$
should be negative that seems to be unphysical because we expect
that in ground state $M=\mu$.
In this connection even the interpretation in terms of antiparticles
does not save a situation because, as was mentioned above, the solution
which could be interpreted as the anti-bell boson state lies in the
rather different parameter space (unlike the $\varphi^4$ theory where
the p-branes with negative $\alpha$ admit physical interpretation).
Therefore, below we will consider the case $c_+$.

As was established in the works \cite{raz,zu}, 
SU(2) has to be the dynamical symmetry
group for this Hamiltonian which can be rewritten in the form of
the spin Hamiltonian
\be
\widehat H= -S^2_z - B S_x,                                 
\ee
where the spin operators,
\ba
&&S_x = S \cosh{\zeta} - \frac{B}{2} \sinh{\!^2 \zeta} - \sinh{\zeta} 
\frac{\drm}{\drm\zeta},  \nn \\
&&S_y = i         \left\{
               -S \sinh{\zeta} + \frac{B}{2} \sinh{\zeta}\cosh{\zeta} + 
\cosh{\zeta} \frac{\drm}{\drm\zeta} \right\},   \\
&&S_z =          \frac{B}{2} \sinh{\zeta}
        + \frac{\drm}{\drm\zeta},              \nn
\ea
satisfy with the commutation relations
\[
[S_i,~S_j] = i \epsilon_{ijk} S_k,                       
\]
besides
\[
S_x^2+S_y^2+S_z^2 \equiv S (S+1).                         
\]
At $S\geq 0$ there exists an irreducible ($2 S+1$)-dimensional 
subspace of the representation space of the su(2) Lie algebra, which is
invariant with respect to these operators.
Determining eigenvalues and eigenvectors of the spin Hamiltonian
in the matrix representation 
which is realized in this subspace, one can prove 
that the solution of (\ref{eq46}) is the function
\ba
\Psi (\zeta) =\exp{
                   \left(
                         -\frac{B}{2} \cosh{\zeta}
                   \right)
                  }
              \sum_{\sigma=-S}^{S}
              \frac{c_\sigma}
                   {
                    \sqrt{
                          (S-\sigma)\verb|!|~
                          (S+\sigma)\verb|!|
                         }
                   }
              \exp{
                   \left(
                         \sigma \zeta
                   \right)
                  }, 
\ea
where the coefficients $c_\sigma$ are the solutions of 
the system of linear equations
\[
\biggl(
       \varepsilon+\sigma^2
\biggr)c_\sigma + \frac{B}{2}
\biggl[
       \sqrt{(S-\sigma)(S+\sigma+1)}~ c_{\sigma+1}             
+ \sqrt{(S+\sigma)(S-
\sigma+1)}~ c_{\sigma-1}
\biggr] = 0,
\]
\[
c_{S+1} = c_{-S-1}=0,~~\sigma=-S,~-S+1,...,~S.            
\]
Regrettably, these expressions give only the 
finite number of exact solutions which is equal to the dimensionality of
the invariant subspace.
Therefore, for the spin $S=0$ we can find only the ground state wave
function and eigenvalue:
\be
\Psi_0 (\zeta) = C_1 
\exp{
    \left(
           - \frac{B}{2} \cosh{\zeta}
    \right)
    },\ 
\varepsilon_0 = 0,
\ee
i.e., we have obtained the expected result that the mass of the 
quantum bell $c_+$-boson
in the ground state coincide with the classical one,
\be                                                         \lb{eq53}
M_0 = \mu.
\ee
In absence of exact wave functions for more excited levels let us find
the first quantum correction to mass of the bell particle
in the approximation of the quantum harmonic oscillator.
It is easy to see that at $B \geq 1$ the (effective) potential 
\be
V(\zeta) = 
\left(
      \frac{B}{2}
\right)^2 \text{sinh}^2 \zeta
-
\frac{B}{2} \cosh{\zeta}
\ee
has the single minimum
\[
V_{\text{min}} = - B/2 \ \ \text{at} \ \ \zeta_{\text{min}}=0.
\]

Then following to the $\hbar$-expansion technique we shift the origin of 
coordinates (to satisfy $\varepsilon = \varepsilon_0 = 0$ in absence
of quantum oscillations) in the point of minimum, and expand $V$ in the 
Taylor series to second order
near the origin thus reducing the model to the oscillator
of the unit mass, energy $\varepsilon/2$ and oscillation frequency
\[
\omega = \frac{1}{2} \sqrt{B(B-1)}.
\]
Therefore, the quantization rules yield the discrete spectrum
\be
\varepsilon =  \sqrt{B(B-1)} (n + 1/2) 
+ O (\hbar^2),\ \ n=0,\ 1,\ 2, ...,
\ee
and the first quantum correction to particle mass will be
determined by the lower energy of oscillations:
\be                                                          \lb{eq56}
\varepsilon = \frac{1}{2} \sqrt{B(B-1)} + O (\hbar^2),
\ee
that gives the algebraic equation for $M$ as a function of 
$m$ and $\lambda$.

We can easily resolve it in the approximation of weak coupling.
Assuming $\lambda \to 0$ (or, equivalently, 
$B \gg 1$) 
we find eq. (\ref{eq56}) in the form
\be                                                         
\varepsilon = \frac{B}{2} + O (\lambda\hbar^2),
\ee
which after considering of eqs. (\ref{eq48}) and (\ref{eq53}) yields
\[
(M-\mu)^2 = \frac{c_+ m^2 M}{8\mu} + O (\lambda\hbar^2),
\]
hence we obtain
\be
M = \mu \pm \frac{1}{2} \sqrt{\frac{c_+}{2}} m + O(\lambda \hbar^2).
\ee
Thus, the mass of the bell boson  with 
quantum corrections at first order of $\hbar$ is
\be                                                      
M = \frac{8 \sqrt{2}}{15} \frac{m^5}{\lambda^2}
\pm
\frac{1}{2}\sqrt{\frac{3}{2}} m,                       
\ee
The quantum term in this expression has to be in good agreement with that
obtained in \cite{dhn} and numerically justified in \cite{kpp}
for the kink solution.
We just point out that the main (classical) term turns to
be singular at $\lambda \to 0$, therefore, the obtained results are 
nonperturbative and can not be ruled out 
from the vacuum sector of the $\varphi^3$ theory 
through the series of the perturbation theory.

\section{Conclusion}\lb{s-c}

It was shown that the cubic scalar field theory admits the 
localized bell-shaped solitary wave interpreted as a 
massive quantum mechanical particle.
This solution has the counterpart of
the same mass which can be imagined as the antiparticle.
The differences in the particle-antiparticle interpretation
from that for the models conserving the parity symmetry 
were studied in this connection.
Further, considering field fluctuations in the neighborhood of 
the static solution, we 
ruled out the p-brane action for the $\varphi^3$-bell 
solutions as the nonminimal point particle with curvature.

When quantizing this action as the constrained
theory with higher derivatives,
it was shown that the resulting \schrod equation has the 
potential from the Razavi class.
This equation
has SU(2) dynamical symmetry group in the ground state
and can be written by virtue of spin operators.
Note, that though the reformulation of some interaction 
concerning the coordinate degrees of freedom in terms of
spin variables is widely used (e.g., in the theories described by
the Heisenberg Hamiltonian, see \cite{lp}), it has to be merely
the physical approximation as a rule,
whereas in our case the spin-coordinate correspondence is exact.
Finally, we found the quantum corrections to the 
mass of the bell bosons which could not be obtained by means
of the perturbation theory starting from the vacuum sector.

\def\CMPh{Commun. Math. Phys.}
\def\JPh{J. Phys.}
\def\CJP{Czech. J. Phys.}
\def\FP{Fortschr. Phys.}
\def\LMPh {Lett. Math. Phys.}
\def\NPh  {Nucl. Phys.}
\def\PhE  {Phys.Essays}
\def\PhL  {Phys. Lett.}
\def\PhR  {Phys. Rev.}
\def\PhRL {Phys. Rev. Lett.}
\def\PhRp {Phys. Rep.}
\def\NCim {Nuovo Cimento}
\def\NuPB {Nucl. Phys.}
\def\GRG {Gen. Relativ. Gravit.}
\def\CQG {Class. Quantum Grav.}
\def\prp {report}
\def\Prp {Report}

\def\jn#1#2#3#4#5{{#1}{#2} {#3} {(#5)} {#4}}   

\def\boo#1#2#3#4#5{ #1 ({#2}, {#3}, {#4}){#5}}  

\def\prpr#1#2#3#4#5{{``#1,''} {#2}{#3}{#4}, {#5} (unpublished)}


\newpage

\begin{thebibliography}{99}

\bibitem{dhn}
R. Dashen, B. Hasslacher, and A. Neveu, 
\jn{\PhR}{ D}{10}{4114}{1974};
\jn{\PhR}{ D}{10}{4130}{1974}; 
\jn{\PhR}{ D}{11}{3434}{1975}. 

\bibitem{kpp}
A. A. Kapustnikov, A. Pashnev, and A. Pichugin, 
\jn{\PhR}{ D}{55}{2257}{1997}. 

\bibitem{col}
S. Coleman, J. Wess, and B. Zumino,
\jn{\PhR}{}{117}{2239}{1969};
C. Callan, S. Coleman, J. Wess, and B. Zumino,
\jn{\PhR}{}{117}{2247}{1969}. 

\bibitem{raj}
R. Rajaraman,
\boo{Solitons and Instantons}{North-Holland}{Amsterdam}{1988}{}.

\bibitem{dhot}
T. Dereli, D. H. Hartley, M. Onder, and R. W. Tucker,
\jn{\PhL}{}{252B}{601}{1990};
M. S. Plyushchay, 
\jn{\NPh}{}{B362}{54}{1991};
V. V. Nesterenko, 
\jn{\CQG}{}{9}{1101}{1992}.

\bibitem{raz} 
M. Razavi,
\jn{Am. \JPh}{}{48}{285}{1980}.

\bibitem{zu} 
V. V. Ulyanov and O. B. Zaslavsky,
\jn{\PhRp}{}{216}{188}{1992};
H. Konwent, P. Machnikowski, and A. Radosz,
\jn{\JPh}{ A}{28}{3757}{1995}.

\bibitem{lp}
Ye. M. Lifshitz and L. P. Pitaevskii,
\boo{Statistical Physics}{Nauka}{Moskow}{1978}{}.

\end{thebibliography}
\end{document}